\documentclass[onecolumn,floatfix,superscriptaddress,a4paper,nofootinbib,notitlepage]{revtex4-2}
\usepackage{epsfig}
\usepackage{latexsym}
\usepackage[dvipsnames]{xcolor}
\usepackage[colorlinks=true,linktocpage=true,linkcolor=blue,citecolor=blue,allcolors=blue]{hyperref}
\usepackage[utf8]{inputenc}
\usepackage{indentfirst}
\usepackage{enumerate}
\usepackage{mathrsfs}
\usepackage[dvipsnames]{xcolor}
\usepackage{eucal}
\usepackage{yfonts}

\usepackage[caption=false,position=top]{subfig}

\usepackage{amsmath}
\usepackage{amssymb}
\usepackage[english]{babel}
\usepackage{url}

\usepackage[section]{placeins}
\usepackage{lineno}

\newcommand{\const}{\mathop{\rm const}\nolimits}

\newcommand{\eq}[1]{\begin{align} #1 \end{align}}

\newcommand{\sNN}{\sqrt{s_{\rm NN}}}

\begin{document}

\title{Hadronic and partonic composition of QCD matter across the crossover}

\author{Artemiy Lysenko}
\affiliation{Physics Department, Taras Shevchenko National University of Kyiv, 03022 Kyiv, Ukraine}
\affiliation{Bogolyubov Institute for Theoretical Physics, 03680 Kyiv, Ukraine}

\author{Mark I. Gorenstein}
\affiliation{Bogolyubov Institute for Theoretical Physics, 03680 Kyiv, Ukraine}

\author{Marek Gazdzicki}
\affiliation{Jan Kochanowski University, 25-369 Kielce, Poland}

\author{Roman Poberezhniuk}
\affiliation{Physics Department, University of Houston, Box 351550, Houston, Texas 77204, USA}
\affiliation{Bogolyubov Institute for Theoretical Physics, 03680 Kyiv, Ukraine}

\author{Volodymyr Vovchenko}
\affiliation{Physics Department, University of Houston, Box 351550, Houston, Texas 77204, USA}

\begin{abstract}
We construct a phenomenological equation of state of strongly interacting matter at zero chemical potentials that provides a unified description of lattice QCD thermodynamics in terms of hadronic and partonic degrees of freedom in the temperature range $T=100$--$510$~MeV. The hadronic phase is described by the quantum van der Waals hadron resonance gas, extended by excluded-volume repulsion between mesons, while the quark--gluon plasma is modeled as an ideal gas of quarks and gluons supplemented with a phenomenological interaction term proportional to $T^3$. The two regimes are connected by a smooth crossover switching function. We determine the four model parameters---the meson hard-core radius, the strength of the partonic interaction term, and two parameters of the switching function---by fitting to lattice QCD results for the trace anomaly. In this model, hadronic contributions remain significant up to about 200~MeV, well above the QCD chiral crossover. We also quantify the crossing contributions generated by the temperature dependence of the switching function. The switching temperature and the hadronic and partonic contributions depend on the chosen equations of state and switching function. The lattice thermodynamic data do not uniquely determine the microscopic composition of QCD matter.
\end{abstract}

\maketitle

\section{Introduction}

Exploring the phase diagram of quantum chromodynamics (QCD), i.e., the structure of phases of strongly interacting matter (SIM) in the plane of temperature $T$ and baryon chemical potential $\mu_B$, is a central theoretical goal of high-energy nuclear physics. Effective models of SIM typically predict a first-order phase transition between hadronic matter and the quark--gluon plasma (QGP) at large baryon chemical potentials. The corresponding transition line in the $\mu_B$--$T$ plane ends at the QCD critical point~\cite{Stephanov:2004wx,Kapusta:2025por}. Available estimates typically place it at $\mu_B \gtrsim 500$~MeV~\cite{Hippert:2023bel,Zhu:2026dmc,Basar:2026irk}. At small chemical potentials, lattice QCD simulations have established that the transition between hadronic matter and the QGP is a smooth crossover~\cite{Aoki:2006we,Bazavov:2011nk,Borsanyi:2020fev}. The early Universe traversed the crossover region during its QCD epoch~\cite{Borsanyi:2016ksw}. Heavy-ion collisions at RHIC and the LHC probe the same region experimentally at the highest available energies, $\sNN \gtrsim 200$~GeV~\cite{Andronic:2017pug,Koch:2025cog,Braun-Munzinger:2025mud}.

Lattice QCD provides first-principles results for the thermodynamic functions at zero chemical potentials~\mbox{\cite{Borsanyi:2013bia,HotQCD:2014kol}}. These data alone, however, do not reveal the underlying microscopic degrees of freedom. Thus, connecting the thermodynamic functions to hadronic and partonic constituents requires modeling. Several approaches have recently addressed this question from different perspectives. One class of approaches constructs phenomenological equations of state that interpolate between hadronic and partonic matter while remaining consistent with lattice QCD thermodynamics~\cite{Albright:2014gva,Kapusta:2025por}.

Another line of research emphasizes that the microscopic composition above the pseudocritical temperature may differ substantially from that of a weakly interacting quark--gluon plasma. In particular, chiral spin symmetry has been interpreted as evidence for an intermediate strongly correlated ``stringy fluid'' regime in which confining chromoelectric interactions remain important~\cite{Glozman:2017dfd,Glozman:2024dzz,Glozman:2025twe}. Complementary studies of quarkyonic matter suggest that confinement may coexist with quark degrees of freedom over a broad range of thermodynamic conditions~\cite{McLerran:2007qj} and have recently explored possible connections of this picture to the finite-temperature QCD phase diagram and lattice simulations~\cite{Fujimoto:2025sxx,McLerran:2026dio}. The present work follows this line of research from a different perspective. Rather than proposing a new microscopic phase of strongly interacting matter, we construct a simple equation of state constrained by lattice QCD thermodynamics and use it to quantify the temperature dependence of the hadronic and partonic contributions across the QCD crossover.

At low temperatures, we describe SIM with the hadron resonance gas (HRG) model with van der Waals interactions among baryons and, separately, among antibaryons~\cite{Vovchenko:2016rkn}, extended here by an excluded-volume repulsion between mesons with a hard-core radius $r_M$. At high temperatures, we model the QGP pressure as an ideal gas of quarks and gluons supplemented with a phenomenological term $-AT^3$, with $A = \const > 0$. The two regimes are connected using the switching function method of Refs.~\cite{Albright:2014gva,Albright:2015uua}. The main differences from the HRG--pQCD constructions of Refs.~\cite{Albright:2014gva,Kapusta:2025por} are the baryonic interactions fixed by nuclear matter properties and the single-parameter phenomenological QGP correction. The parameter $k$ controls the shape of the switching function, while $T_0$, which we call the switching temperature, sets its temperature scale. We determine the four parameters, $r_M$, $A$, $k$, and $T_0$, by fitting the lattice data on the trace anomaly, $(\varepsilon - 3p)/T^4$. We then compare the resulting equation of state (EoS) with the lattice pressure, entropy density, energy density, and speed of sound at $T=100$--$510$~MeV. Since these quantities follow from the same pressure, their agreement with the lattice data is a consistency check of the fit. The hadronic contributions remain significant up to $T\approx200$~MeV, well above the pseudocritical temperature of the chiral transition, $T_{\rm pc}\approx155$~MeV~\cite{Borsanyi:2010bp,HotQCD:2014kol}, and the chemical freeze-out temperature, $T_{\rm ch}=150$--$160$~MeV~\cite{Andronic:2017pug,Vovchenko:2015cbk}.

The paper is organized as follows. We formulate the HRG model with the meson excluded volume in Sec.~\ref{HRG}. We present the QGP model in Sec.~\ref{QGP}. We construct the full EoS based on the switching function in Sec.~\ref{SFFEoS}, where we fix the model parameters by fitting the lattice data. Section~\ref{Sum} summarizes the results and discusses their microscopic interpretation.

\section{Hadron resonance gas}
\label{HRG}

To describe the hadronic phase of SIM, we employ the quantum van der Waals hadron resonance gas (QvdW-HRG) model proposed in Ref.~\cite{Vovchenko:2016rkn}. Unlike the ideal HRG, this model incorporates non-resonant attractive and repulsive baryon interactions in a thermodynamically consistent manner while preserving quantum statistics. It reproduces the liquid--gas phase transition and ground-state properties of nuclear matter, providing a useful phenomenological description of hadronic matter.

Following Ref.~\cite{Vovchenko:2016rkn}, we include van der Waals interactions only among baryons and, separately, among antibaryons. The original formulation neglects baryon--antibaryon, meson--meson, and meson--(anti)baryon van der Waals interactions. The pressure in the grand canonical ensemble reads
\begin{equation}
\label{p-H}
p_H(T,\mu)=P_M+P_B+P_{\bar B}~,
\end{equation}
where the meson and (anti)baryon partial pressures are
\eq{\label{P-MX}
P_M=\sum_{i\in M}p_i^{\rm id}(T,\mu_i)~, \qquad
P_X=\sum_{i\in X}p_i^{\rm id}(T,\mu_i^{X*})-a\,n_X^2~, \quad
X\in\{B,\bar B\}~.
}
Here $M$ denotes the set of all mesons, while $B$ and $\bar{B}$ denote the sets of baryons and antibaryons, respectively. The chemical potentials $\mu=(\mu_B,\mu_Q,\mu_S)$ are conjugate to the three conserved charges of strong interactions: baryon number, electric charge, and strangeness. The chemical potential of the $i$-th hadron species is $\mu_i=b_i\mu_B+q_i\mu_Q+s_i\mu_S$, where $b_i$, $q_i$, and $s_i$ denote the corresponding conserved charges.

The ideal-gas pressure of the $i$-th hadron species is
\begin{equation}
\label{p-id}
p_i^{\rm id}(T,\mu_i)=
\frac{d_i}{6\pi^2}
\int dm\,f_i(m)
\int_0^\infty
\frac{dk\,k^4}{\sqrt{k^2+m^2}}
\left[
\exp\left(
\frac{\sqrt{k^2+m^2}-\mu_i}{T}
\right)
\pm1
\right]^{-1},
\end{equation}
where the plus sign corresponds to fermions and the minus sign to bosons. Here $d_i$ is the spin degeneracy factor, while the mass integration over the distribution $f_i(m)$ accounts for finite resonance widths, with $f_i(m)\rightarrow\delta(m-m_i)$ for stable hadrons. We treat the resonance widths using the energy-independent Breit--Wigner prescription~\cite{Vovchenko:2018fmh}.  We use the PDG2020 hadron list implemented in the \texttt{Thermal-FIST}   package~\cite{Vovchenko:2019pjl}, which contains all established hadrons and resonances listed in the 2020 edition of the Particle Data Group. We do not include light nuclei in the list. 

The shifted chemical potentials $\mu_i^{X*}$ of (anti)baryons are determined from the transcendental equation
\eq{\label{mu-X}
\mu_i^{X*}
=
\mu_i
-
b\,P_X
-
ab\,n_X^2
+
2an_X~,
}
which accounts simultaneously for the repulsive excluded-volume correction and the attractive mean-field contribution. The (anti)baryon number densities satisfy
\begin{equation}
\label{n-X}
n_X
=
(1-bn_X)
\sum_{i\in X}
n_i^{\rm id}(T,\mu_i^{X*})~,
\end{equation}
where
$n_i^{\rm id}(T,\mu_i)
=
(\partial p_i^{\rm id}/\partial\mu_i)_T$
is the ideal-gas number density.

The parameters $a>0$ and $b>0$ characterize, respectively, the mean-field attractive interaction and the excluded-volume repulsion between (anti)baryons. We fix the parameters by reproducing the empirical saturation properties of symmetric nuclear matter at zero temperature~\cite{Vovchenko:2015pya}. The resulting values, $a=329~{\rm MeV\,fm}^3$ and $b=3.42~{\rm fm}^3$, are assumed to be universal for all baryon and antibaryon species~\cite{Vovchenko:2016rkn}.

\begin{figure}[h!]
\centering
\includegraphics[width=.49\textwidth]{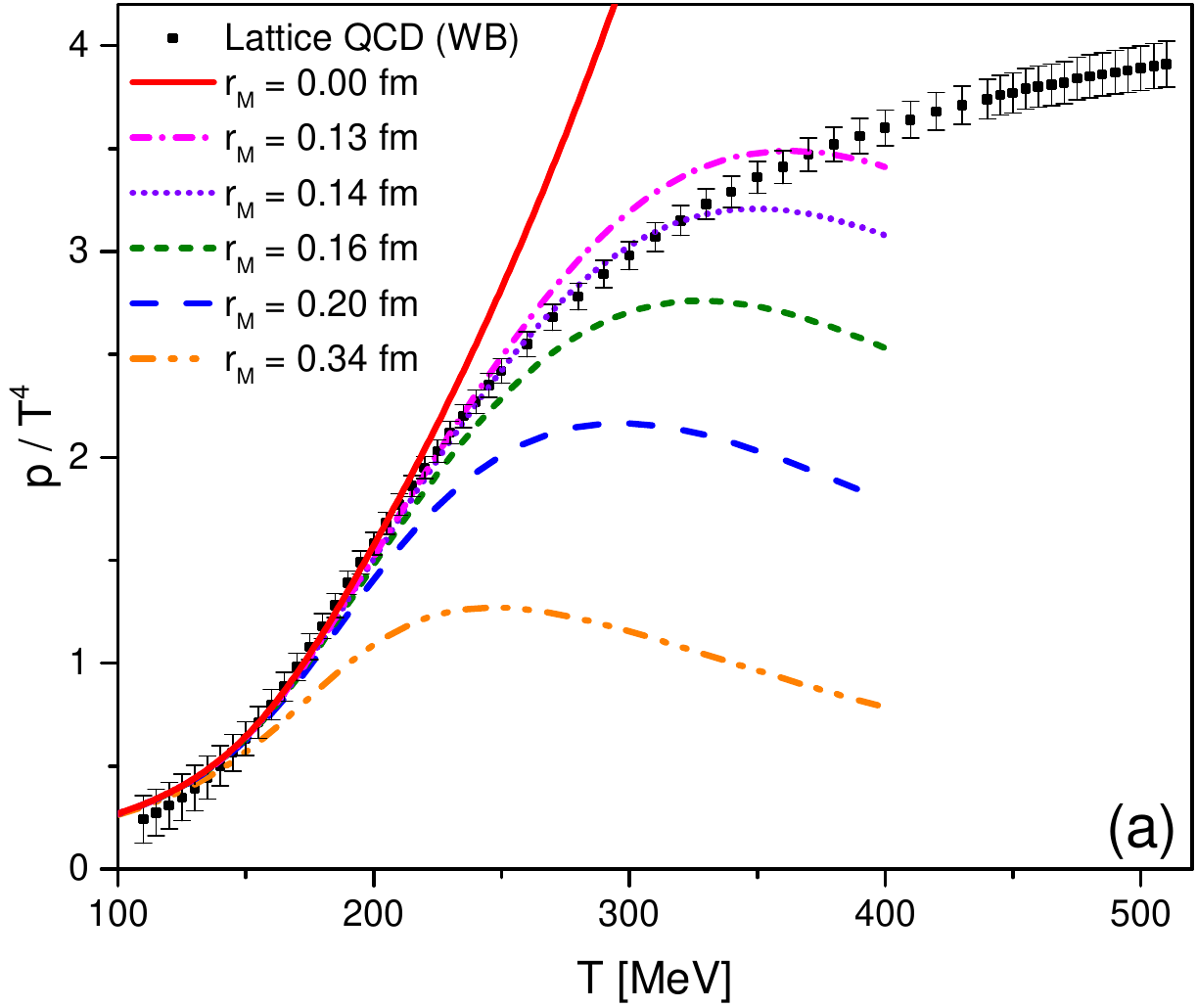}
\includegraphics[width=.49\textwidth]{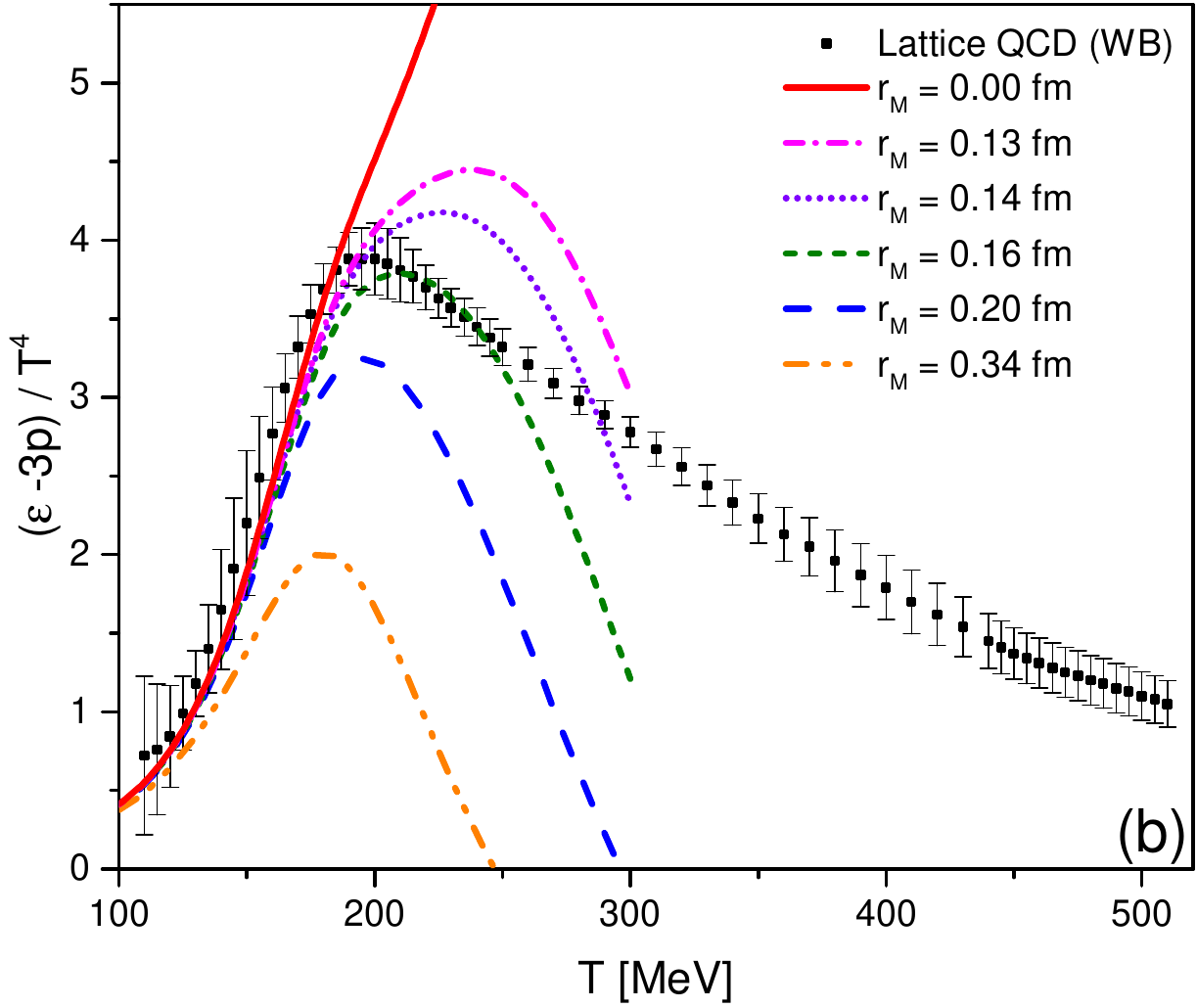}
\caption{Temperature dependence of (a) the scaled pressure $p/T^4$ and (b) the trace anomaly $(\varepsilon-3p)/T^4$ in the QvdW-HRG model at $\mu=0$ for several values of the meson hard-core radius $r_M$. The case $r_M=0$ corresponds to the original model of Ref.~\cite{Vovchenko:2016rkn} with point-like mesons. We use lattice QCD data from the Wuppertal--Budapest Collaboration (Ref.~\cite{Borsanyi:2013bia}).}
\label{figHRG}
\end{figure}

The QvdW-HRG pressure at vanishing chemical potentials, $\mu=0$, corresponds to the $r_M=0$ curve in Fig.~\ref{figHRG}(a). We perform all numerical calculations using the \texttt{Thermal-FIST} package~\cite{Vovchenko:2019pjl}. The model reproduces lattice QCD thermodynamics\footnote{We use the Wuppertal--Budapest (WB) data \cite{Borsanyi:2013bia} as they span a broader temperature range ($100$--$510$ MeV) compared to the HotQCD data \cite{HotQCD:2014kol}. More accurate lattice data for (2+1+1)-flavor matter have been presented recently~\cite{Borsanyi:2025dyp}. They are consistent within errors with the data of Ref.~\cite{Borsanyi:2013bia} that we use.}~\cite{Borsanyi:2013bia} up to temperatures of about $T\lesssim200$ MeV, but predicts a much steeper increase of the pressure at higher temperatures. Such behavior prevents a thermodynamically consistent matching to the deconfined phase because, if the HRG and QGP represent different phases of SIM, the Gibbs criterion requires the QGP pressure to exceed the HRG pressure at sufficiently high temperatures~\cite{Landau:1980mil}.

Including excluded-volume repulsion only for baryons is insufficient because, at high temperatures, point-like mesons dominate the thermodynamics, and their effective degrees of freedom exceed those of the QGP. This indicates that the point-like meson assumption becomes inadequate at high particle densities. We therefore extend the QvdW-HRG model by introducing short-range excluded-volume repulsion among mesons. This modification is not intended as a microscopic description of meson interactions. Rather, it provides a simple phenomenological parametrization of the reduction of the effective hadronic degrees of freedom at high temperatures while preserving the successful description of low-temperature thermodynamics. This description has been applied in several studies to suppress hadrons at high temperatures and densities, particularly in the context of building a hadron-quark description of QCD thermodynamics~\cite{Steinheimer:2010ib,Vovchenko:2014pka,Motornenko:2019arp}. The meson chemical potentials in Eq.~(\ref{P-MX}) are replaced according to
\eq{\label{mu-M}
\mu_i
\rightarrow
\mu_i^{M*}
=
\mu_i
-
b_M
\sum_{j\in M}
p_j^{\rm id}(T,\mu_j^{M*})~,
}
where, unlike in the baryonic sector, we include only the excluded-volume correction and introduce no attractive mean-field contribution. The meson number density satisfies
\begin{equation}
\label{n-M}
n_M
=
(1-b_Mn_M)
\sum_{i\in M}
n_i^{\rm id}(T,\mu_i^{M*})~.
\end{equation}
The excluded-volume parameter is related to the meson hard-core radius by
$b_M=\frac{16}{3}\pi r_M^3$.

In contrast, while nucleon--nucleon interactions are well constrained in nuclear physics, little empirical information is available on short-range repulsion between mesons, although interacting meson matter has been described within the van der Waals equation of state~\cite{Poberezhnyuk:2015dba,Redlich:2016dpb}. Furthermore, the attractive meson--meson and meson--(anti)baryon interactions are expected to be largely encoded in the resonance spectrum included in the HRG~\cite{Dashen:1969ep}. Because experimental constraints on short-range mesonic repulsion remain limited, we treat the meson hard-core radius $r_M$ as a phenomenological free parameter. Figure~\ref{figHRG}(a) illustrates the temperature dependence of the HRG pressure for several values of $r_M$. Even a modest meson excluded volume suppresses the rapid increase in HRG pressure at high temperatures, ensuring that the QGP pressure eventually dominates.

At vanishing chemical potentials, the entropy density and energy density follow from the thermodynamic identities
\eq{\label{iden}
s=\frac{dp}{dT}~,~~~~
\varepsilon=T\frac{dp}{dT}-p~.
}
Figure~\ref{figHRG}(b) shows the corresponding trace anomaly, $(\varepsilon-3p)/T^4$, for the same values of $r_M$.
Unlike the original QvdW-HRG model with point-like mesons, finite $r_M$ values produce a pronounced maximum in the trace anomaly. For $r_M\simeq0.16$ fm, the HRG model alone reproduces the position and magnitude of the lattice QCD peak around $T\simeq200$ MeV. This underscores that a pronounced maximum in the trace anomaly is not, by itself, a unique signature of deconfined matter but may also arise from strong hadronic interactions~\cite{Asakawa:1995zu,Satz:2011wf,Xu:2012kc}.
This hadron-only choice of $r_M$ is illustrative; the fit of the full EoS in Sec.~\ref{SFFEoS} determines $r_M\cong0.34$~fm together with $k$, $T_0$, and $A$.

\section{Quark-gluon plasma}
\label{QGP}

To describe the QGP phase of strongly interacting matter, we start from an ideal gas of quarks and gluons and introduce a simple phenomenological correction that parametrizes interaction effects remaining above the QCD transition. The resulting parametrization provides a minimal, thermodynamically consistent description of the lattice QCD equation of state while keeping the number of free parameters as small as possible.

The pressure of an ideal gas of quarks and gluons at $\mu=0$ is
\begin{equation}
\label{PQ-id}
p_Q^{\rm id}(T)
=
\frac{d_g\pi^2}{90}T^4
+
\sum_{q=u,d,s}
\frac{d_q}{6\pi^2}
\int_0^\infty
\frac{dk\,k^4}
{\sqrt{k^2+m_q^2}}
\left[
\exp\left(
\frac{\sqrt{k^2+m_q^2}}{T}
\right)
+1
\right]^{-1},
\end{equation}
where $d_g=16$ and $d_q=12$ are the numbers of internal degrees of freedom of gluons and of each quark flavor (including antiquarks), respectively. We include the three lightest quark flavors, $u$, $d$, and $s$, consistent with the $N_f=2+1$ lattice QCD calculations~\cite{Borsanyi:2013bia} analyzed below.

In the considered temperature range, the bare masses of the $u$ and $d$ quarks can be neglected, yielding
\begin{equation}
\label{p-ud}
p^{\rm id}_{u/d}(T)
=
\frac78
\frac{d_q\pi^2}{90}
T^4~.
\end{equation}
We take the bare strange-quark mass from the Particle Data Tables: $m_s\simeq93.5$ MeV~\cite{ParticleDataGroup:2024cfk}.

\begin{figure}[h!]
\centering
\includegraphics[width=.49\textwidth]{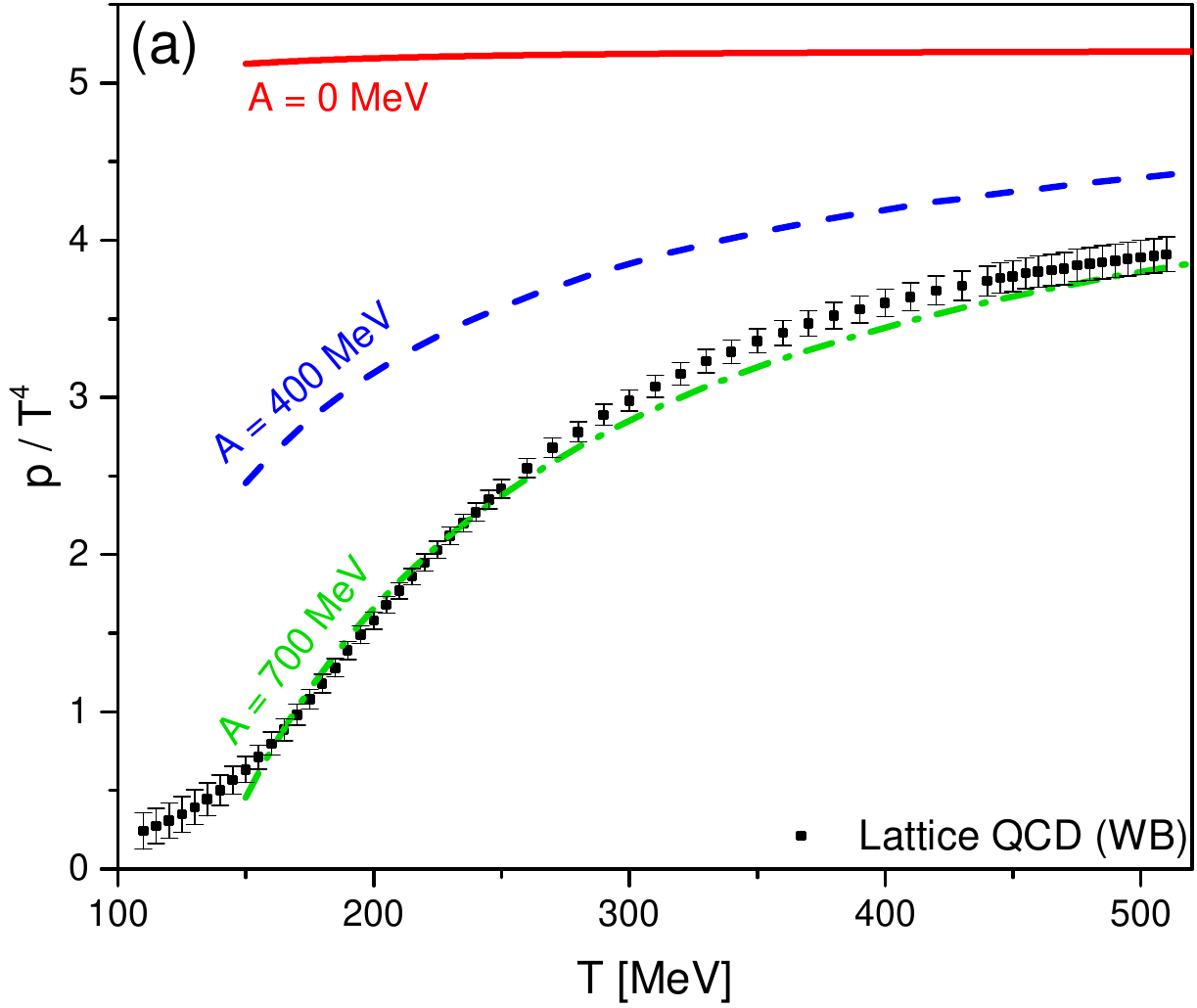}
\includegraphics[width=.49\textwidth]{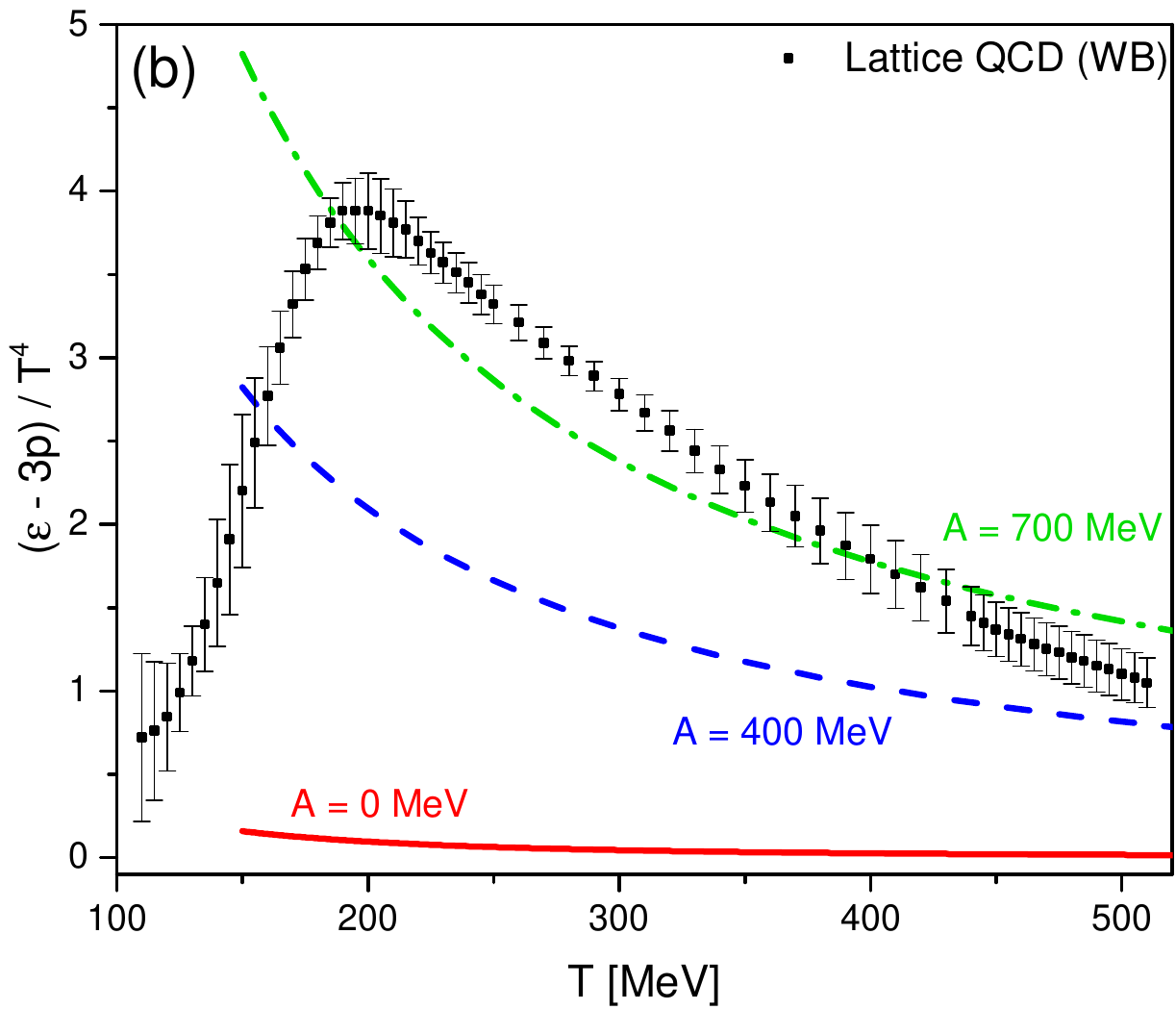}
\caption{Temperature dependence of (a) the scaled pressure $p/T^4$ and (b) the trace anomaly $(\varepsilon-3p)/T^4$ in the QGP model~(\ref{pQ}) for several values of the interaction parameter $A$. The case $A=0$ corresponds to the ideal quark--gluon gas~(\ref{PQ-id}). We use lattice QCD data from the Wuppertal--Budapest Collaboration in Ref.~\cite{Borsanyi:2013bia}.}
\label{figQGP}
\end{figure}

The quantity $p_Q^{\rm id}/T^4$ corresponds to the $A=0$ curve in Fig.~\ref{figQGP}(a). It approaches the Stefan--Boltzmann limit of a weakly interacting quark--gluon plasma,
\[
\frac{p}{T^4}
=
\frac{\pi^2}{90}
\left(
d_g+\frac{21}{8}d_q
\right)
\simeq5.2~,
\]
the small deviation originates from the finite strange-quark mass. In contrast, the lattice QCD pressure remains significantly below this limit even at the highest temperature considered, $T=510$ MeV, indicating that sizable interaction effects persist well above the crossover region.

To account for these interaction effects while preserving the simplicity of the model, we supplement the ideal-gas pressure by a phenomenological interaction term,
\begin{equation}
\label{pQ}
p_Q(T)
=
p_Q^{\rm id}(T)
-
AT^3,
\end{equation}
where $A>0$ is a temperature-independent parameter with the dimension of energy. The correction to $p_Q/T^4$ decreases as $1/T$ relative to the dominant ideal-gas contribution, and therefore vanishes asymptotically, ensuring that the Stefan--Boltzmann limit is recovered at sufficiently high temperatures.

Figure~\ref{figQGP}(a) shows $p_Q/T^4$ for several values of $A$. For $A=700$ MeV, the model reasonably describes the lattice pressure above $T\simeq200$ MeV. Later, by fitting the combined hadron-quark model to the lattice data, we obtain a somewhat smaller value, $A\simeq624$~MeV, which reflects the residual contribution of the hadronic phase in the crossover region. We calculate the entropy density and the energy density from the thermodynamic identities~(\ref{iden}). Figure~\ref{figQGP}(b) shows the resulting trace anomaly.

\newpage

Neglecting the small strange-quark mass correction, Eq.~(\ref{pQ}) yields
\[
\frac{\varepsilon_Q-3p_Q}{T^4}
\simeq
\frac{A}{T},
\]
which decreases monotonically with temperature. Consequently, the QGP model (\ref{pQ}) alone cannot reproduce the pronounced peak of the trace anomaly observed in lattice QCD. As shown in the next section, this peak emerges naturally from matching the QGP and HRG equations of state. We determine the parameter $A$ and the meson hard-core radius $r_M$ simultaneously by fitting the full equation of state to lattice QCD results.

\section{Unified equation of state}
\label{SFFEoS}

\begin{figure}[h!]
\centering
\includegraphics[width=.485\textwidth]{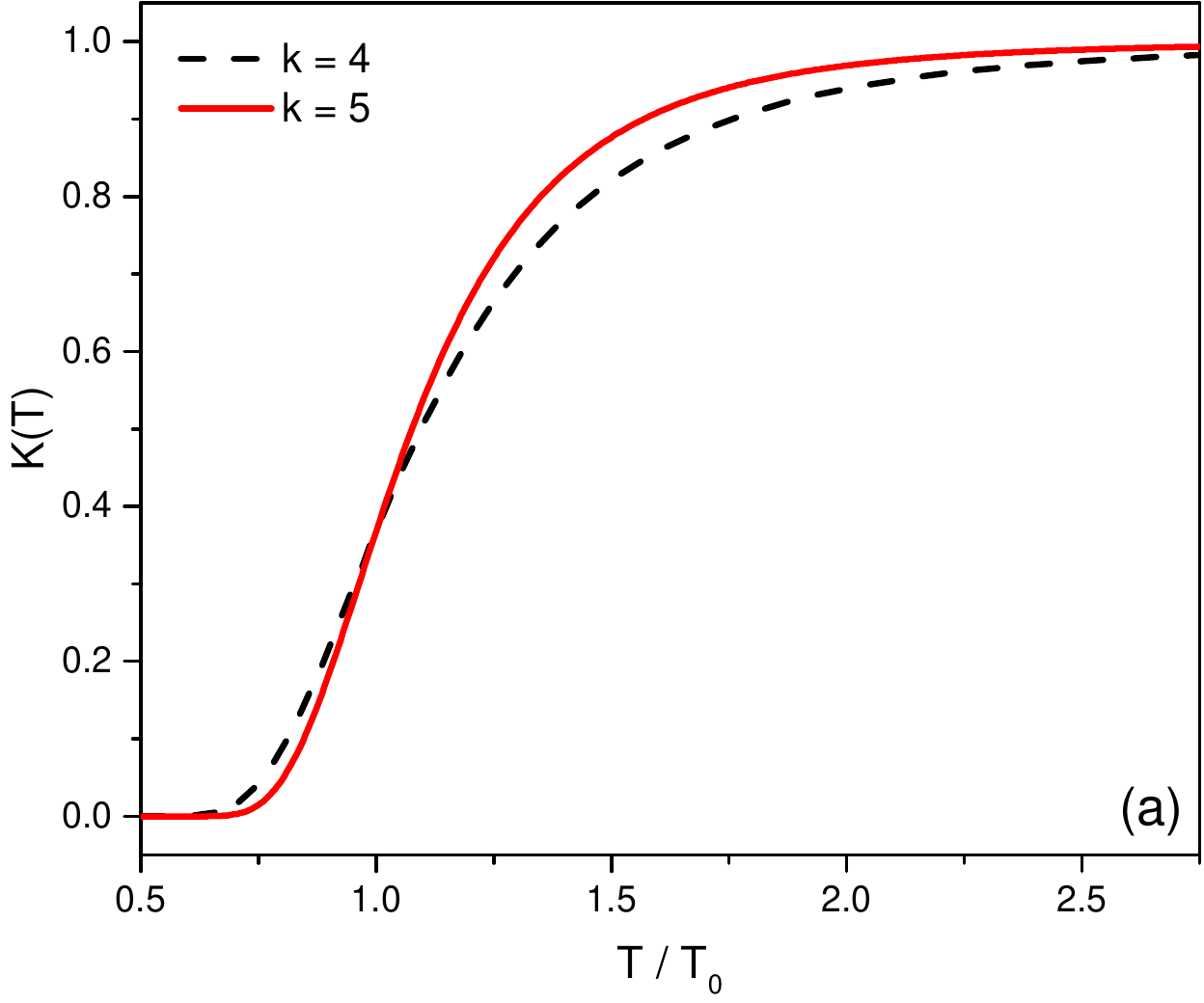}
\includegraphics[width=.49\textwidth]{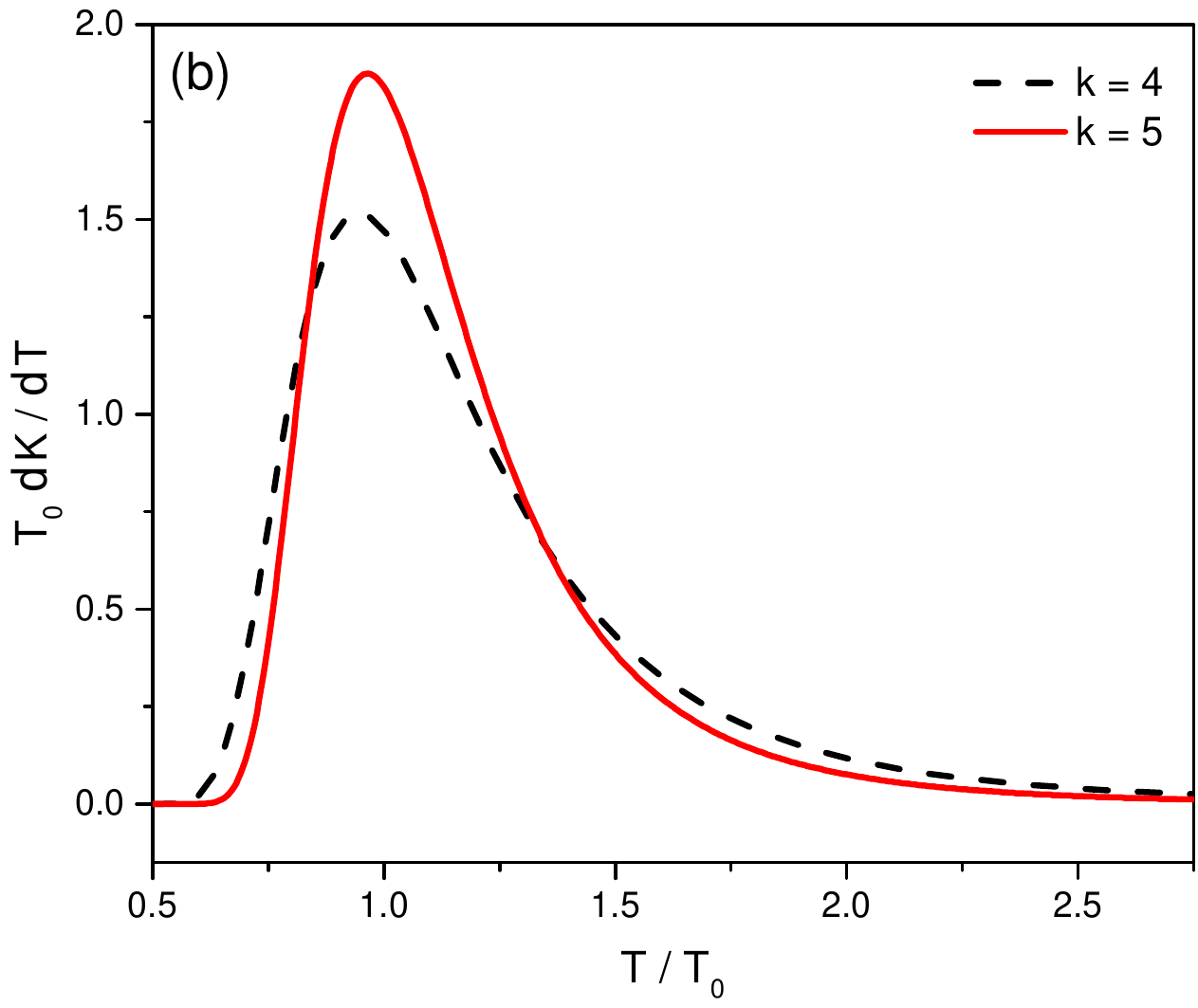}
\caption{(a) The switching function $\mathscr{K}$ and (b) its derivative $T_0\,d\mathscr{K}/dT$ for $k=4$ and 5 as functions of $T/T_0$, which corresponds to the best fits (see below).}
\label{switching_function}
\end{figure}

To describe the smooth crossover between the hadronic and quark--gluon plasma phases, we employ the switching-function method proposed in Ref.~\cite{Albright:2014gva}. Unlike the Maxwell construction appropriate for a first-order phase transition, the switching-function approach provides a thermodynamically smooth interpolation between the HRG and QGP equations of state without introducing discontinuities in thermodynamic observables. At vanishing chemical potentials, the full equation of state of strongly interacting matter is written as
\begin{equation}
\label{p-full}
p(T)=\mathscr{K}(T)\,p_Q(T)+\left[1-\mathscr{K}(T)\right]p_H(T),
\end{equation}
where $p_H$~(\ref{p-H}) and $p_Q$~(\ref{pQ}) denote the HRG and QGP pressures, respectively. The switching function $\mathscr{K}$ and its temperature derivative are taken as
\begin{equation}
\label{K-switch}
\mathscr{K}(T)=
\exp\left[-\left(\frac{T_0}{T}\right)^k\right]~,~~~~\frac{d \mathscr{K}(T)}{d T} =
\frac{k}{T}
\left(\frac{T_0}{T}\right)^k
\mathscr{K}(T)~.
\end{equation}
Following Ref.~\cite{Albright:2014gva}, we use Eq.~(\ref{K-switch}) for the switching function with positive integer $k$. This function is infinitely differentiable and introduces no non-analytical features in the considered EoS (\ref{p-full}). Figure~\ref{switching_function} shows the switching function and its $T$-derivative as functions of $T/T_0$ for $k=4$ and 5. The function $\mathscr{K}$ rapidly approaches zero at $T\ll T_0$ and unity at $T\gg T_0$. The weights $\mathscr{K}$ and $1-\mathscr{K}$ multiply the QGP and HRG pressures, respectively, in Eq.~(\ref{p-full}). The parameter $k$ regulates the shape of $\mathscr{K}$, which approaches the theta function $\theta(T/T_0-1)$ at $k\gg1$. The switching temperature $T_0$ sets the interpolation temperature scale, with $\mathscr{K}(T_0)=e^{-1}\cong0.37$ for any $k$. The pressure weights become equal at $T_{1/2}=T_0(\ln2)^{-1/k}$, and $d\mathscr{K}/dT$ reaches its maximum at $T^*=T_0[k/(k+1)]^{1/k}$. For $k=5$, these temperatures are $T_{1/2}\cong1.08~T_0$ and $T^*\cong0.96~T_0$.

Interpolating the pressure is particularly convenient because all remaining thermodynamic observables follow uniquely from standard thermodynamic identities, ensuring internal thermodynamic consistency. The entropy density is obtained from Eq.~(\ref{iden}),
\begin{equation}
\label{s-full}
s=\frac{dp}{dT}
=
\mathscr{K}(T)\,s_Q
+
\left[1-\mathscr{K}(T)\right]s_H
+
\frac{d \mathscr{K}(T)}{d T} \left(p_Q-p_H\right)~\equiv~\mathscr{K}s_Q+(1-\mathscr{K})s_H+
s_\times,
\end{equation}
where $s_Q=dp_Q/dT$, $s_H=dp_H/dT$ are calculated with the QGP and HRG EoS, respectively, and $\mathscr{K}$ and $d\mathscr{K}/dT$ are given by Eq.~(\ref{K-switch}). The temperature dependence of $\mathscr{K}$ generates the crossing term $s_\times$ in the entropy density and hence $Ts_\times$ in the energy density and trace anomaly,
\begin{align}
& \varepsilon~\equiv~Ts-p ~=~ \mathscr{K}\varepsilon_Q~+~ (1-\mathscr{K}) \varepsilon_H ~+~Ts_{\times}~,
\label{epsilon}\\
 &\varepsilon-3p~ =~ \mathscr{K}(\varepsilon_Q -3p_Q) ~ +~ (1-\mathscr{K})(\varepsilon_H -3p_H)~+~Ts_{\times}~.
 \label{trace}
\end{align}
Thus, $\mathscr{K}$ and $1-\mathscr{K}$ are pressure weights, not fractions of QGP and hadronic matter.

\begin{figure}[h!]
\centering
\includegraphics[width=.49\textwidth]{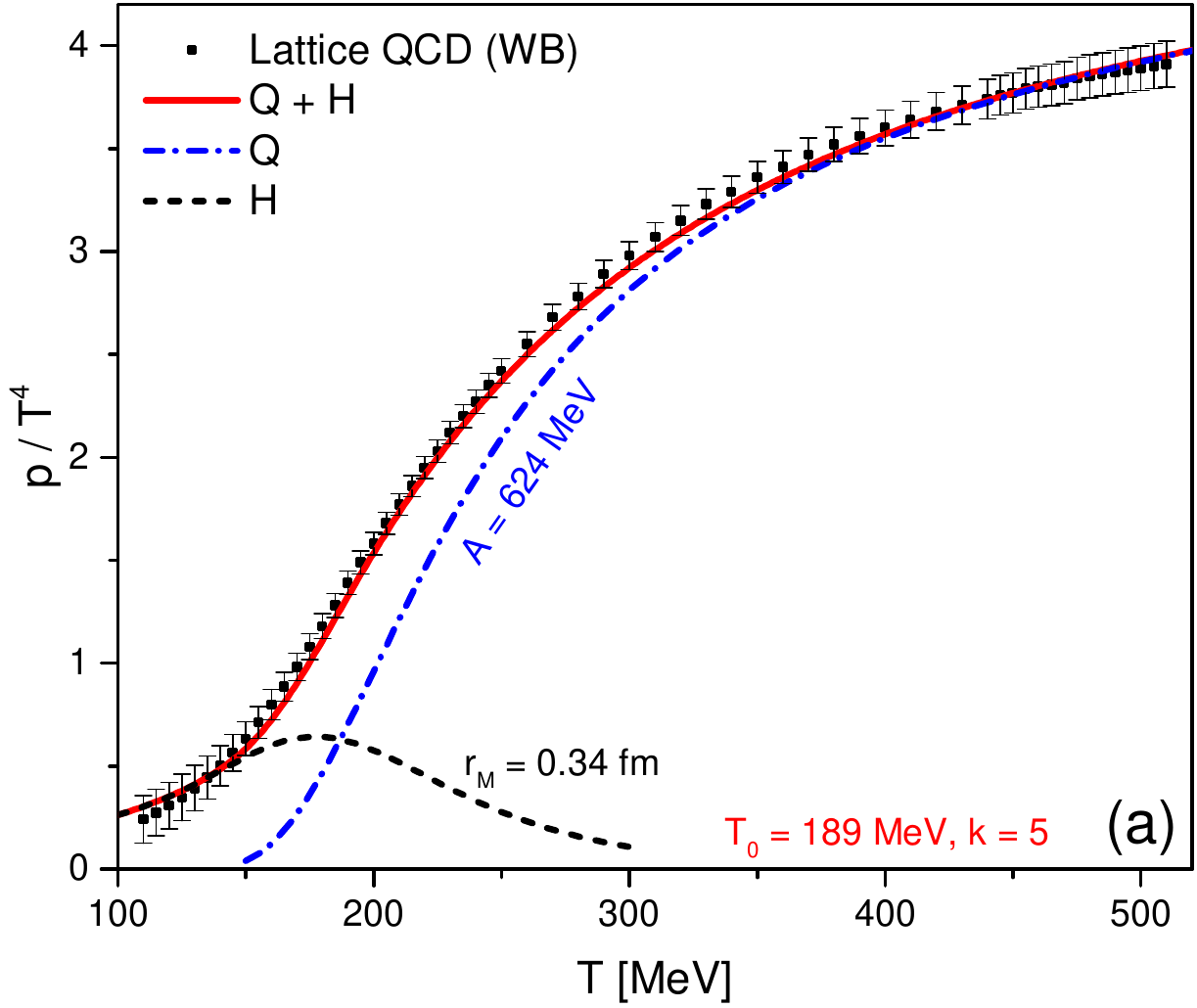}
\includegraphics[width=.49\textwidth]{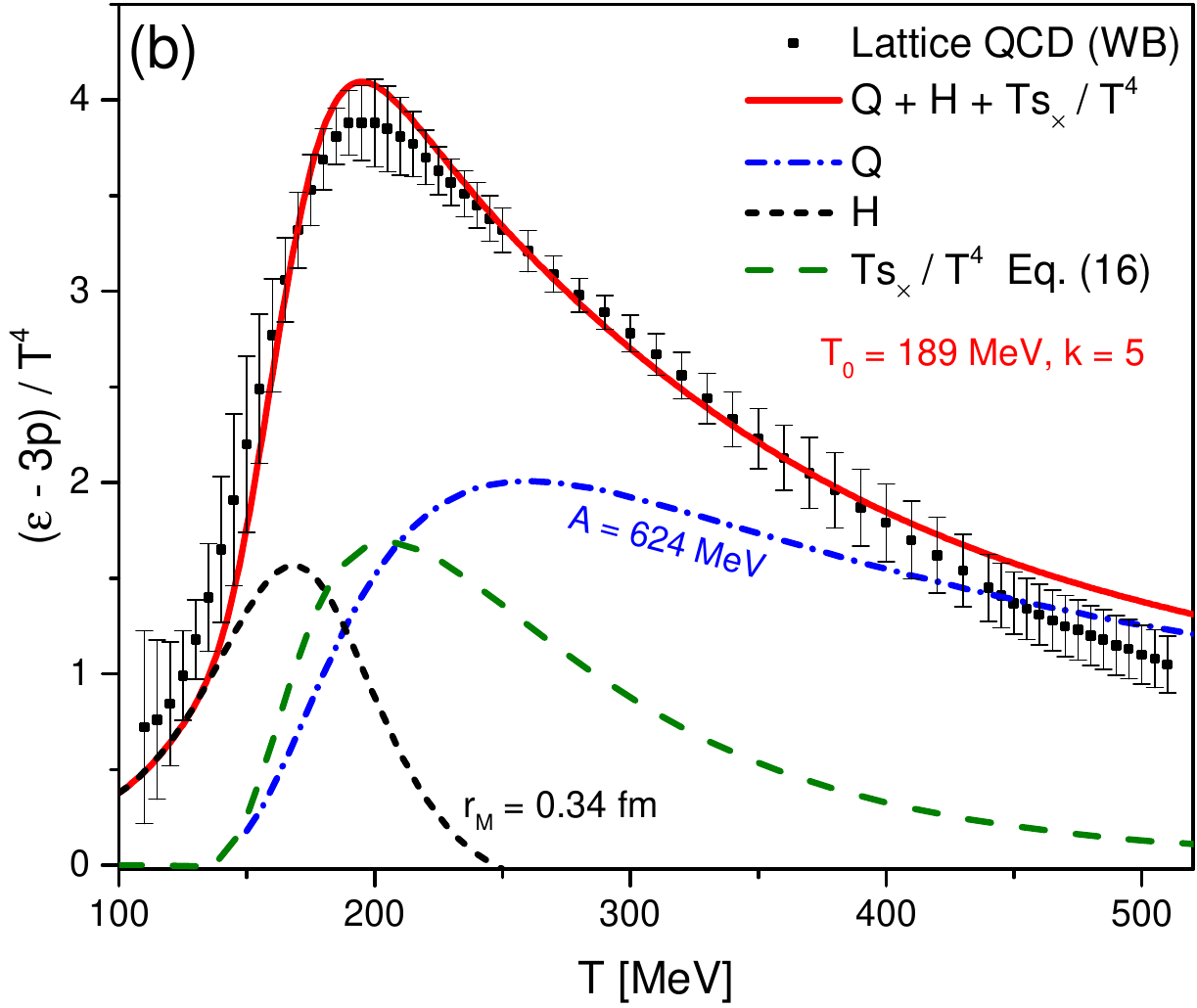}
\caption{Temperature dependence of (a) the scaled pressure $p/T^4$ and (b) the trace anomaly $(\varepsilon-3p)/T^4$ for the full equation of state (\ref{p-full}). The dash-dotted and short-dashed curves show the QGP and HRG terms of Eqs.~(\ref{p-full}) and~(\ref{trace}), and the long-dashed curve in (b) shows the crossing term. We take the lattice QCD data of the Wuppertal--Budapest Collaboration from Ref.~\cite{Borsanyi:2013bia}.}
\label{figFull}
\end{figure}

\begin{figure}[t]
\centering
\includegraphics[width=.32\textwidth]{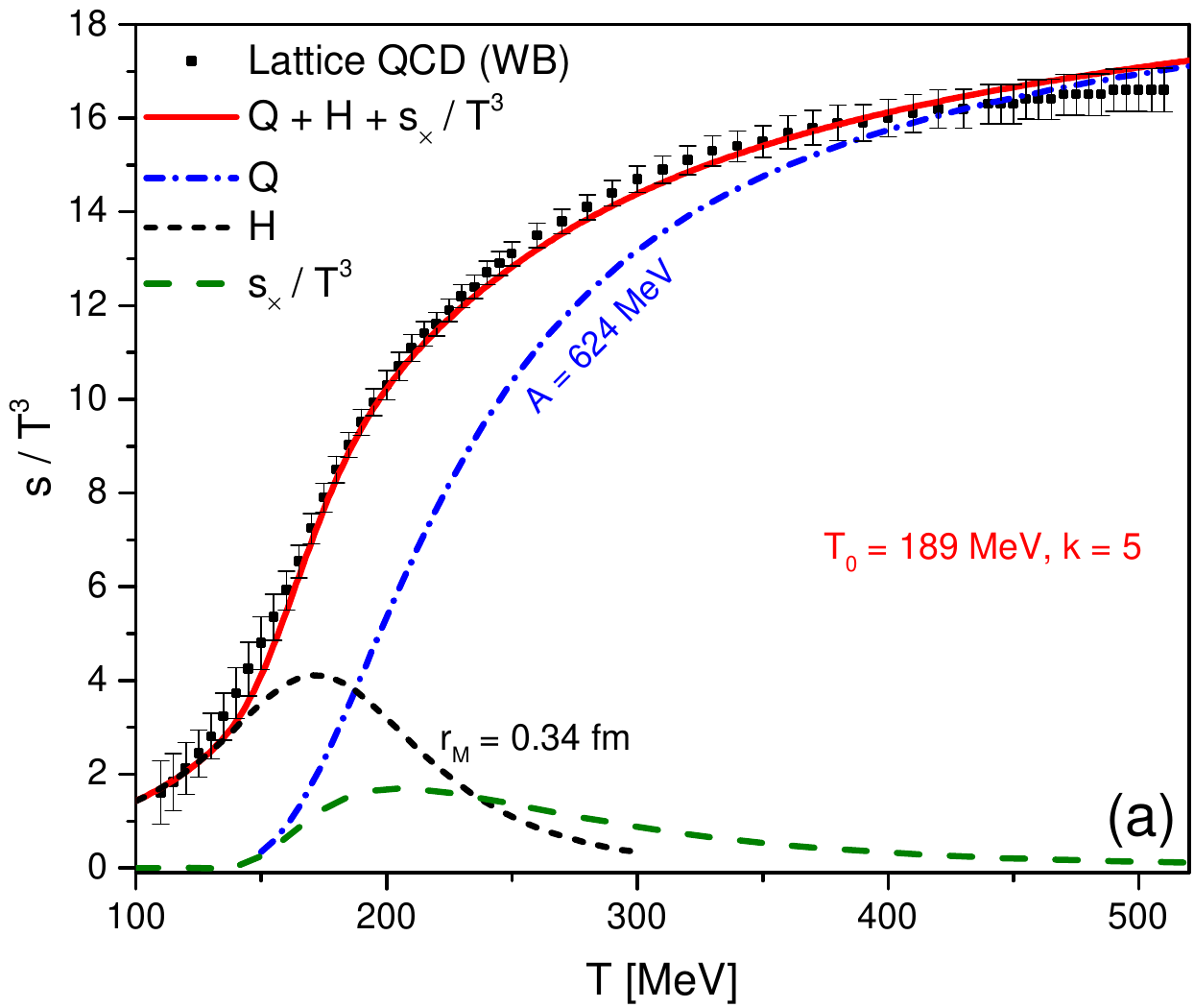}
\includegraphics[width=.32\textwidth]{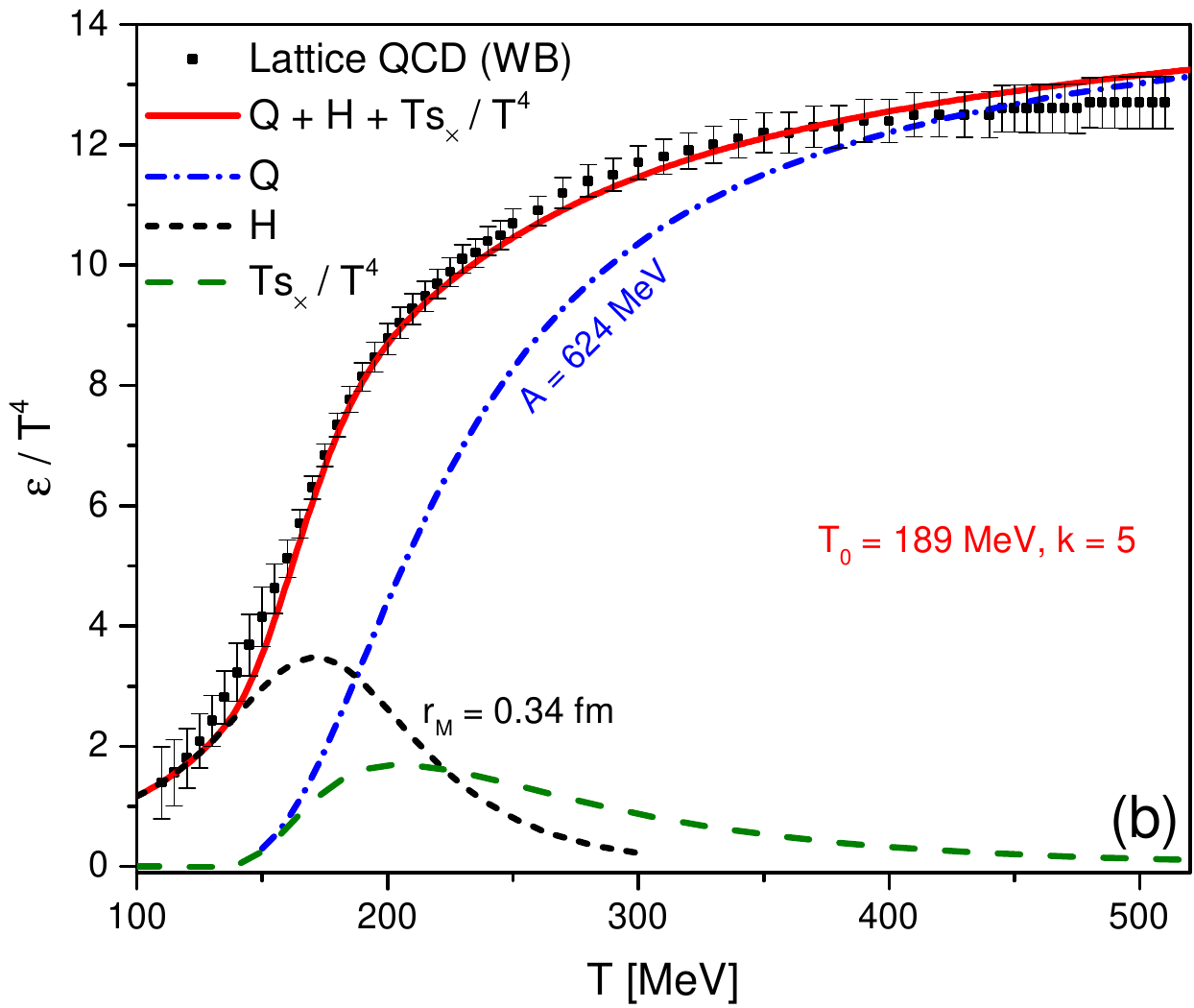}
\includegraphics[width=.33\textwidth]{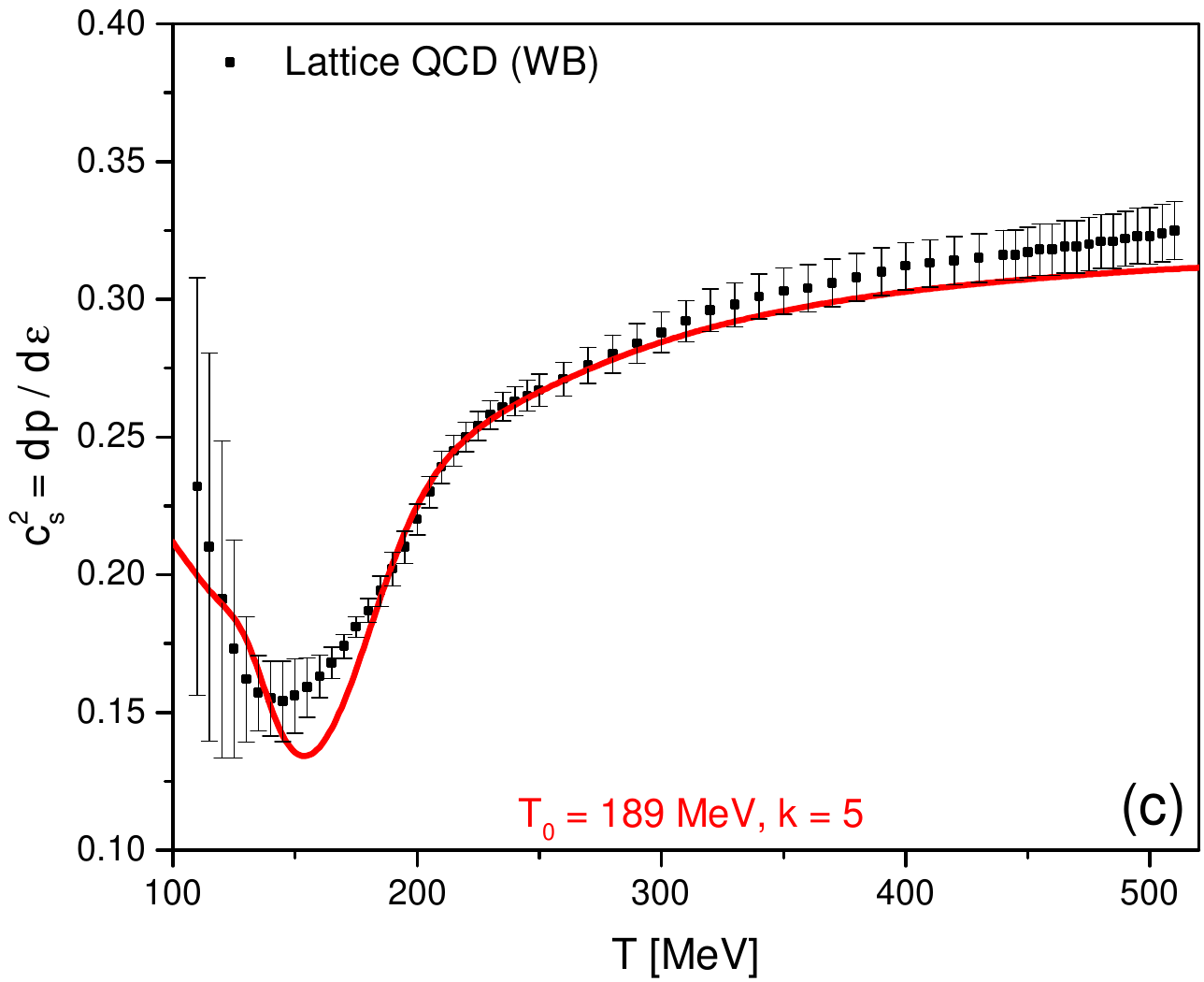}
\caption{Temperature dependence of (a) the entropy density $s/T^3$, (b) the energy density $\varepsilon/T^4$, and (c) the speed of sound squared, $c_s^2=dp/d\varepsilon$, for the full equation of state (\ref{p-full}). In (a) and (b), the dash-dotted, short-dashed, and long-dashed curves show the QGP, HRG, and crossing terms of Eqs.~(\ref{s-full}) and~(\ref{epsilon}), respectively. We use the lattice QCD data from the Wuppertal--Budapest Collaboration in Ref.~\cite{Borsanyi:2013bia}.}
\label{figSE}
\end{figure}

Near $T\approx200$~MeV, the crossing terms account for about $16\%$ of the entropy density, $19\%$ of the energy density, and $41\%$ of the trace anomaly (Figs.~\ref{figFull} and~\ref{figSE}). The crossing contributions in Eqs. (\ref{s-full})-(\ref{trace}) are proportional to $d \mathscr{K}/dT$, vanish at $T\ll T_0$ and $T\gg T_0$ (see Fig. \ref{switching_function}(b)). We note that the exact analytical form of the switching function remains an open question and may warrant further study~\cite {Monnai:2019hkn,Miyahara:2019zfn}. For instance, a recent statistical-mixture formulation~\cite{Yang:2026brr} proposes that the mixing weight becomes an internal thermodynamic variable determined by minimizing the grand potential.

Our present analysis is restricted to zero chemical potentials. Extending it to finite chemical potentials is a possible future extension, which will require modifications. These modifications include both formulating the QGP phase EoS at finite $\mu_B$ and the possible explicit dependence of the switching function $\mathscr{K}$ on $\mu_B$~(or any other chemical potentials).  In turn, the derivatives of $\mathscr{K}$ with respect to $\mu_B$ will contribute to the baryon density and higher-order baryon-number susceptibilities. Lattice QCD provides the susceptibilities at $\mu_B = 0$. However, they can receive contributions from the chemical-potential dependence of both the switching function and the QGP interaction term, neither of which is constrained by thermodynamics at vanishing chemical potentials. We leave such an extension to finite chemical potentials for future work.

Our complete EoS (\ref{p-full}) of QCD matter at vanishing chemical potentials contains four phenomenological parameters: the meson hard-core radius $r_M$, the QGP interaction parameter $A$, and the two parameters of the switching function $k$ and $T_0$. We determine them from a $\chi^2$ fit to lattice data for the trace anomaly $(\varepsilon - 3p)/T^4$. Table~\ref{tabI} presents the results.

\begin{table}[h!]
    \centering
    \begin{tabular}{| c | c | c | c | c |}
\hline
~$k$~  & $T_0$ [MeV] & $A$ [MeV] & $r_M$ [fm] & $\chi^2/ \rm NDF$ \\
\hline
$3$ & $182 \pm 2$ & $448 \pm 1$ & $0.66 \pm 0.06$ & $1.34$ \\
$4$ & $179 \pm 4$ & $587 \pm 1$ & $0.46 \pm 0.04$ & $1.04$ \\
$5$ & $189 \pm 4$ & $624 \pm 1$ & $0.34 \pm 0.03$ & $1.04$ \\
$6$ & $190 \pm 4$ & $652 \pm 2$ & $0.31 \pm 0.03$ & $1.37$ \\
\hline
\end{tabular}
    \caption{Values of the parameters of the full EoS (\ref{p-full}) obtained by fitting the lattice QCD data of the Wuppertal--Budapest Collaboration \cite{Borsanyi:2013bia} for the trace anomaly $(\varepsilon - 3p)/T^4$.}
    \label{tabI}
\end{table}

\vspace{0.2cm}

Our best fits correspond to $k=4$ and $k=5$. Although both $k=4$ and $k=5$ produce an identical $\chi^2/ \rm NDF$ of $1.04$ and represent our best fits, for certainty we prefer to select $k=5$ as the primary value for all subsequent calculations. The choice of $k=5$ is also due to the smaller value of the meson radius (Table~\ref{tabI}), which leads to smaller changes in the EoS of the HRG at $T=160$ MeV. The model parameters presented in Table~\ref{tabI} allow to establish the broader acceptable uncertainty ranges (\ref{wide}) for the model parameters in Sec.~\ref{Sum}. 

We use $k=5$ together with
\begin{equation}\label{best}
T_0~=~189~{\rm MeV}~,~~~A~=~624~{\rm MeV}~,~~~r_M~=~0.34~{\rm fm}~,
\end{equation}
in all further calculations. The quoted $\chi^2/{\rm NDF}$ is calculated using the published uncertainties shown in Ref.~\cite{Borsanyi:2013bia}, which correspond to the diagonal elements of the covariance matrix. The lattice QCD results at different temperatures are known to be strongly correlated, but the full covariance matrix is not publicly available. Therefore, the quoted $\chi^2/ \rm NDF$ in Table~\ref{tabI} should be regarded primarily as a measure of the overall quality of the description rather than as a statistically rigorous goodness-of-fit estimator. We then obtain all remaining thermodynamic observables from the thermodynamic identities (\ref{iden}).

Figure~\ref{figFull} shows the resulting pressure and trace anomaly, together with the QGP and HRG terms of Eq.~(\ref{p-full}) in panel (a) and the three terms of Eq.~(\ref{trace}) in panel (b). The HRG, QGP, and crossing terms all contribute to the trace-anomaly peak near $T\approx200$~MeV.

Figure~\ref{figSE} compares the full equation of state with the lattice QCD results for the entropy density, $s/T^3$, the energy density, $\varepsilon/T^4$, and the speed of sound squared, $c_s^2=dp/d\varepsilon$. The model reproduces the lattice data within uncertainties over nearly the entire temperature range. These quantities follow from the pressure~(\ref{p-full}) through the thermodynamic identities~(\ref{iden}), so their agreement with the lattice data is a consistency check of the fit, not an independent test. The speed of sound is a ratio of temperature derivatives and does not separate into HRG, QGP, and crossing terms (Appendix~\ref{app-speed-of-sound}), so Fig.~\ref{figSE}(c) shows only the full result.

Let us now discuss the physical interpretation of the model parameters, $r_M$, $A$, and $T_0$.

\bigskip

\noindent
{\bf The hard-core meson radius $r_M$.} 
The (anti)baryon excluded-volume parameter $b=3.42~\mathrm{fm}^3$, corresponding to a baryon hard-core radius $r_B\cong0.59$~fm, is fixed by nuclear matter properties and is not refitted. We introduce $r_M>0$ because point-like mesons otherwise dominate the HRG at high temperature and drive its pressure above the QGP pressure. We continue to neglect meson--baryon cross terms and attractive nonresonant meson interactions. The value $r_M\simeq0.16$~fm in Fig.~\ref{figHRG}(b) illustrates how the HRG alone can reproduce the trace-anomaly peak. The full-EoS fit~(\ref{best}) instead gives $r_M\simeq0.34$~fm, corresponding to $b_M=16\pi r_M^3/3\cong0.66$~fm$^3$. This is considerably smaller than the (anti)baryon excluded volume, $b=3.42$~fm$^3$. Consequently, meson excluded-volume effects are expected to have only a small influence at low temperatures. Indeed, Fig.~\ref{figHRG}(a) shows that the meson excluded volume with $r_M=0.34$~fm reduces the HRG pressure only slightly at $T\lesssim160$~MeV. Therefore, meson excluded-volume effects are expected to have little influence on thermal fits of hadron multiplicities at the chemical freeze-out temperature, $T_{\rm ch}=150$--$160$~MeV~\cite{Andronic:2017pug,Vovchenko:2015cbk,Poberezhnyuk:2019pxs}, and other standard applications of the HRG model. In contrast, at $T\gtrsim200$~MeV mesonic excluded-volume effects substantially suppress the HRG pressure. The obtained value of the meson hard-core radius is comparable to those studied in the literature in various contexts~\cite{Vovchenko:2014pka,Motornenko:2020yme,Kuznietsov:2021lax,Yang:2026brr}.

\bigskip

\noindent
{\bf The phenomenological interaction term $-AT^3$.}
The behavior of the thermodynamic functions in lattice QCD indicates interaction effects in the QGP even at the highest temperature considered, $T = 510$~MeV. The literature offers different explanations for these effects. In Refs.~\cite{Albright:2014gva,Kapusta:2025por} the quark-gluon interactions were described by perturbative correction terms up to order $\alpha_s^3$, with two free parameters adjusted to fit the lattice results. In our QGP EoS model, we instead describe all interaction effects with the phenomenological term $-AT^3$, which contains a single free parameter. The fitted value, $A\simeq624$~MeV, corresponds to substantial interaction effects: at $T = 200$~MeV, the QGP pressure~$p_Q$ and energy density $\varepsilon_Q$ are reduced by about $60\%$ and $40\%$, respectively, compared to their ideal gas values. At $T=200$~MeV, the QGP cannot be considered as a weakly interacting quark-gluon gas, and the applicability of the purely perturbative description of the QGP in the temperature range $150$--$250$~MeV therefore remains questionable. The parameter $A$ is of the order of the QCD scale,
\[
A
\simeq
1.9\,
\Lambda^{N_f=3}_{\overline{\rm MS}}~,
\]
using the value reported in Ref.~\cite{FlavourLatticeAveragingGroupFLAG:2021npn}.

Power-law corrections to the QGP pressure have been discussed before.
We compare the forms
\begin{equation}
 p_Q^{(n)}(T)=p_Q^{\mathrm{id}}(T)
 ~-~ B_nT^{n}~,~~~~n=0,1,2,3~,
 \label{p-B}
\end{equation}
where $B_n>0$ is a constant. The MIT bag model~\cite{Chodos:1974je} corresponds to $n=0$. A linear term, $-B_1T$, suggested in Ref.~\cite{Gorenstein:1988fe}, suppresses the pressure but leaves the energy density unchanged. The fuzzy-bag picture and dimension-two condensates suggest a quadratic term, $-B_2T^2$~\cite{Pisarski:2006hz,Pisarski:2006yk,Megias:2009mp}. Our ansatz~(\ref{pQ}) corresponds to $n=3$ with $B_3\equiv A$. In Appendix~\ref{app-powerlaw}, we compare these forms with the lattice pressure over a wide temperature range and show that only $n=3$ reproduces the temperature dependence of the lattice pressure. This comparison supports the $-AT^3$ term in the temperature range studied. The term is phenomenological and is not derived from QCD. Changing the temperature dependence, for example by reducing the coefficient of $T^4$ and adding lower powers of $T$, or by using perturbative corrections~\cite{Albright:2014gva,Kapusta:2025por}, would change the fitted parameters. The value of $A$, of order $\Lambda_{\rm QCD}$, is qualitatively consistent with the picture in which non-perturbative confining correlations survive in the QGP well above the crossover~\cite{Glozman:2017dfd,McLerran:2007qj}. For $T \lesssim 360$~MeV the interaction term $AT^3$ exceeds the ideal-gas pressure of gluons, $d_g \pi^2 T^4/90$. The QGP pressure~(\ref{pQ}) in this temperature range therefore stays below that of an ideal gas of quarks alone. This is consistent with the expectation that gluonic degrees of freedom are strongly suppressed just above the crossover, whether because thermal gluon masses exceed the quark ones~\cite{Peshier:1994zf,Levai:1997yx} or because gluons remain confined in heavy glueballs up to about $285$~MeV~\cite{Fujimoto:2025sxx}.

Effective thermal masses of quarks and gluons provide another way to suppress $p_Q^{\rm id}$~\cite{Peshier:1994zf,Gorenstein:1995vm,Peshier:1995ty,Levai:1997yx,Vovchenko:2018eod}. We do not pursue this approach here because it requires additional parameters.

\bigskip

\noindent
{\bf The switching temperature $T_0$.} The pseudocritical temperature of the chiral crossover at $\mu_B=0$, $T_{\rm pc}\approx155$~MeV~\cite{Aoki:2006we,Borsanyi:2010bp,HotQCD:2014kol}, is close to the chemical freeze-out temperature, $T_{\rm ch}=150$--$160$~MeV. The fitted switching temperature, $T_0\simeq189$~MeV, is considerably higher and agrees with Ref.~\cite{Albright:2014gva}. A similar picture emerges in the chiral mean-field model~\cite{Motornenko:2019arp,Motornenko:2020vqm}, where the hadronic and quark contributions to the total pressure become equal at a comparable temperature~\cite{Motornenko:2020vqm}. For the fitted parameters, the hadronic pressure weight in Eq.~(\ref{p-full}) decreases from $1-\mathscr K\simeq0.93$ at $T=155$~MeV to $0.22$ at $T=250$~MeV. Because of the crossing terms in Eqs.~(\ref{s-full})--(\ref{trace}), this weight is not the hadronic fraction of the entropy or energy density. The hadronic contributions shown in Figs.~\ref{figFull} and~\ref{figSE} remain significant up to $T\approx200$~MeV. It will be interesting to investigate alternative forms of the switching function~\cite{Kapusta:2025por,Miyahara:2019zfn} and the statistical-mixture approach proposed in Ref.~\cite{Yang:2026brr}, and to incorporate conserved-charge susceptibilities into the analysis.

\section{Summary and discussion}
\label{Sum}

We constructed a simple phenomenological equation of state for strongly interacting matter (SIM) at vanishing chemical potentials, relevant for describing relativistic heavy-ion collisions at the RHIC and LHC. The hadronic EoS is described by the quantum van der Waals hadron resonance gas (QvdW-HRG) model~\cite{Vovchenko:2016rkn}, extended by introducing an excluded-volume repulsion between mesons. We model the deconfined phase as a gas of quarks and gluons supplemented by a phenomenological interaction term equal to $-AT^3$. The switching-function method~\cite{Albright:2014gva} connects the two limiting descriptions smoothly and provides a thermodynamically consistent realization of the crossover transition at $\mu_B=0$.

The complete equation of state contains four phenomenological parameters. Our choise of the model parameters obtained from the fit of  the lattice QCD trace anomaly gives the meson hard-core radius, $r_M\simeq0.34$~fm, the QGP interaction parameter, $A\simeq624$~MeV, and the switching-function parameters $k=5$ and $T_0\simeq189$~MeV. The resulting equation of state describes the lattice QCD thermodynamic functions over the temperature range $T=100$--$510$~MeV. Because these functions follow from the same pressure, this agreement is a consistency check of the fit rather than an independent test. The fit with $k=4$ gives the same $\chi^2/{\rm NDF}$ value close to unity as for $k=5$ (see Table~\ref{tabI}). Including these fits we obtain the broader parameter ranges
\begin{equation}\label{wide}
T_0~=~175\text{--}195~{\rm MeV}~,~~~A~=~585\text{--}625~{\rm MeV}~,~~~r_M~=~0.3\text{--}0.5~{\rm fm}~,
\end{equation}
in addition to the best fit~(\ref{best}). Lattice data on conserved-charge fluctuations at zero chemical potentials could further constrain these parameters.

The proposed model leads to four physical conclusions. First, the peak of the trace anomaly is not by itself a unique signature of deconfined matter. It can be reproduced within the hadronic sector alone once short-range repulsive interactions between mesons are included.  For the full EoS, it comes from a combination of the HRG, QGP, and the crossing-term contributions from the switching function. Second, the phenomenological nonperturbative term $-AT^3$ gives a significant correction to the ideal-gas QGP pressure well above the crossover region. Third, the switching temperature, $T_0\simeq189$~MeV, is considerably higher than both the pseudocritical chiral-crossover temperature, $T_{\rm pc}\approx155$~MeV, and the chemical freeze-out temperature, $T_{\rm ch}=150$--$160$~MeV. Consequently, the hadronic contributions to the EoS remain significant well above $T_{\rm pc}$, up to $T\approx200$~MeV. Fourth, the crossing contributions to the entropy density, energy density, and trace anomaly arising in the transition region, $T=150$--$250$~MeV, are significant. They are proportional to the derivative of the switching function. The switching temperature and the hadronic, partonic, and crossing contributions depend on the chosen HRG and QGP equations of state and switching function. Our conclusions about the composition of QCD matter are therefore model dependent.

In addition to providing an accurate parametrization of lattice QCD thermodynamics, the proposed equation of state provides additional information not directly accessible from lattice calculations alone. Examples include the full strangeness content, i.e., the sum of strange plus antistrange charges inside the SIM, and the possibility of evaluating quantities relevant for dynamical simulations of heavy-ion collisions, such as thermal photon and dilepton emission. These physical quantities cannot be calculated from lattice QCD thermodynamics data, but depend on the system's degrees of freedom. The constructed model can therefore be useful for heavy-ion phenomenology at RHIC and the LHC. 

The present results also raise several open questions. The sizable hadronic terms of our EoS up to temperatures of about 200~MeV suggest that the properties of hadronic matter in the crossover region may differ qualitatively from those of the hadron resonance gas near chemical freeze-out. This observation is consistent with recent ideas that color-singlet hadronic excitations or hadron-like correlations may survive well above the pseudocritical temperature and coexist with deconfined quark and gluon degrees of freedom in the crossover region~\mbox{\cite{Glozman:2017dfd,Glozman:2024dzz,Glozman:2026glk,McLerran:2007qj,McLerran:2026dio}}. Interpreting the interval $150$--$200$~MeV as a distinct phase of strongly interacting matter would, however, raise new questions about the nature of the two crossover transitions bounding it, near $T\simeq150$~MeV and $T\simeq200$~MeV. Our phenomenological equation of state does not introduce a third phase.

The present work also has several limitations that indicate natural directions for future studies. First, the phenomenological high-temperature equation of state should be confronted with perturbative QCD descriptions~\cite{Albright:2014gva,Kapusta:2025por} to clarify which features of the lattice thermodynamics genuinely require additional nonperturbative contributions. Second, a global fit including both hadronic and high-temperature parameters, together with alternative switching functions or statistical-mixture formulations, would provide an important test of the robustness of the extracted switching temperature. Finally, extending the analysis to finite baryon chemical potential and incorporating conserved-charge susceptibilities will allow one to investigate whether the picture obtained here remains valid away from $\mu_B=0$ and whether it can provide additional constraints on the structure of strongly interacting matter.

\begin{acknowledgments}
We are thankful to Musfer Adzhymambetov, Leonid Glozman, Joseph Kapusta, Volodymyr Kuznietsov, Larry McLerran, Robert Pisarski,  Lorenzo Salcedo, and Jan Steinheimer,  for fruitful comments and discussions. M.I.G. and A.L. thank the National Research Foundation of Ukraine for support under grant number 2025.07/0050.
\end{acknowledgments}

\newpage
\appendix
\section{Speed of sound}
\label{app-speed-of-sound}

The speed of sound squared follows from the full EoS~(\ref{p-full}) as a ratio of temperature derivatives,
\begin{equation} \label{app_2_3}
    c^2_s = \frac{dp}{d \varepsilon} = \frac{dp/dT}{d \varepsilon/dT} = \frac{\mathscr{K}s_Q + (1-\mathscr{K})s_H + s_\times}{\mathscr{K}\frac{d \varepsilon_Q}{d T} + (1-\mathscr{K})\frac{d \varepsilon_H}{d T} + \frac{d \mathscr{K}}{d T}(\varepsilon_Q - \varepsilon_H) + T\frac{d s_\times}{d T} + s_\times}~.
\end{equation}
The denominator, $d\varepsilon/dT=T\,d^2p/dT^2$, contains the second temperature derivatives of the pressure and of the switching function. Unlike the entropy density, energy density, and trace anomaly in Eqs.~(\ref{s-full})--(\ref{trace}), $c_s^2$ is therefore not a sum of HRG, QGP, and crossing terms. For this reason, Fig.~\ref{figSE}(c) shows only the full result~(\ref{app_2_3}).

\section{Power-law corrections to the QGP pressure}
\label{app-powerlaw}

\begin{figure}[h!]
\centering
\includegraphics[width=.70\textwidth]{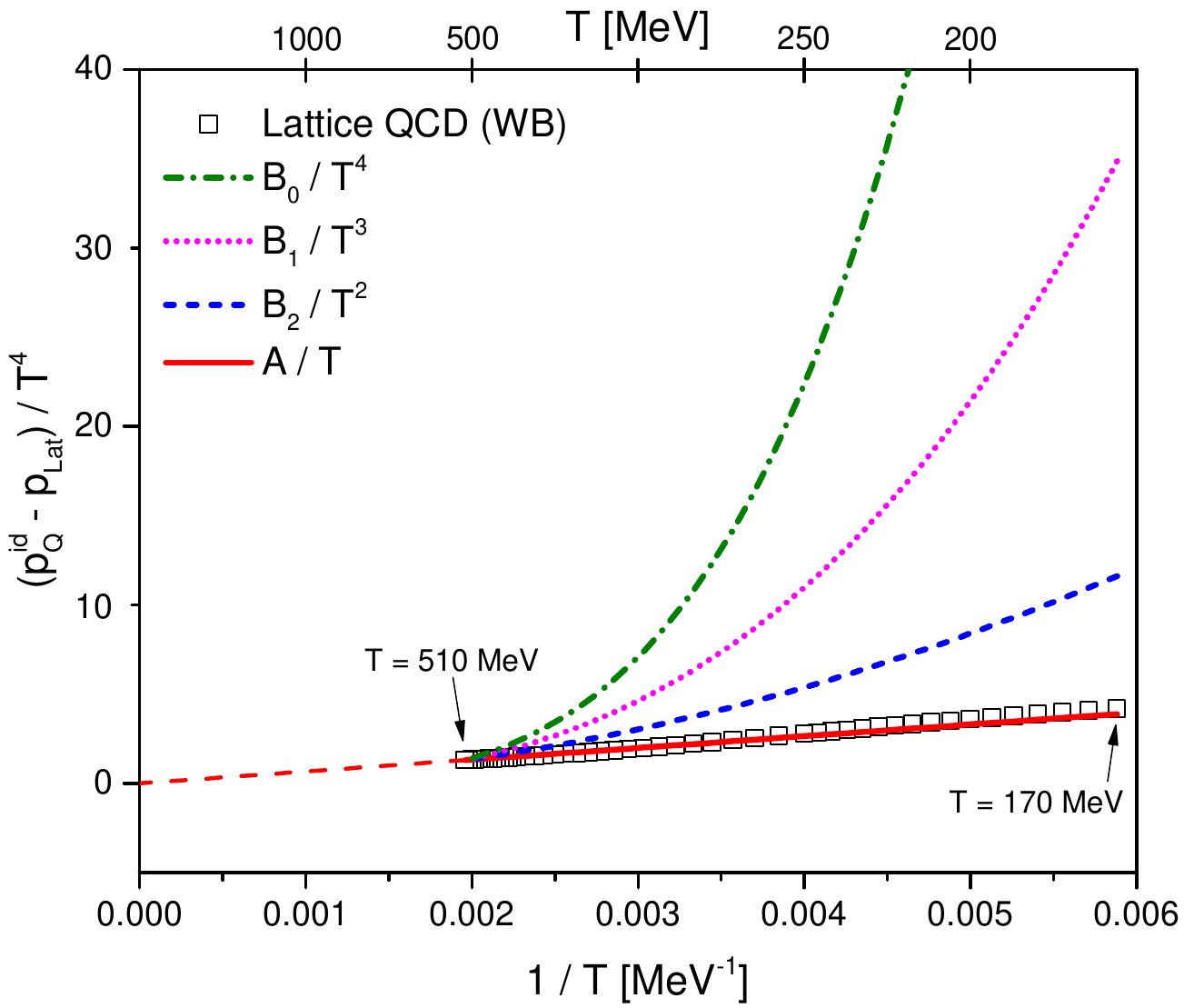}
\caption{Difference between the ideal QGP pressure~(\ref{PQ-id}) and the lattice QCD pressure of the Wuppertal--Budapest Collaboration~\cite{Borsanyi:2013bia}, scaled by $T^4$, as a function of $1/T$ (open squares). The lines show the corrections $B_nT^{n-4}$ of Eq.~(\ref{p-B}) for $n=0$ (dash-dotted), $n=1$ (dotted), $n=2$ (dashed), and $n=3$ (solid; $B_3/T\equiv A/T$ in the legend), each matched to the lattice pressure at $T=510$~MeV. The red dashed line extends the $n=3$ line to $1/T\rightarrow0$.}
\label{fig-B}
\end{figure}

Here we compare the power-law corrections~(\ref{p-B}) with the lattice QCD pressure $p_{\rm Lat}$~\cite{Borsanyi:2013bia}. Figure~\ref{fig-B} shows the difference between the ideal QGP pressure and the lattice pressure, scaled by $T^4$, as a function of $1/T$ (open squares), together with the power-law forms of Eq.~(\ref{p-B}), for which
\begin{equation}
\frac{p_Q^{\rm id}-p_{\rm Lat}}{T^4}~=~B_n\,T^{n-4}~.
\label{p-diff}
\end{equation}
We fix each $B_n$ by matching to the lattice pressure at the highest available temperature, $T=510$~MeV. Therefore, the value of the constant $B_3 \equiv A$ in Fig. \ref{fig-B} differs from the value of parameter $A$ obtained from the lattice data approximation and indicated in Table \ref{tabI}.

The lattice points follow the straight line $B_3/T \equiv A/T$. The forms with $n=0$, 1, and 2 deviate strongly from the lattice points. This comparison clearly favors the $-AT^3$ term in the equation of state for the quark-gluon plasma (\ref{pQ}); this option also remains preferable for the pressure $p_Q$ in the full equation of state (\ref{p-full}).

\bibliography{bibliography}

\end{document}